\documentclass[twocolumn,tighten,dvipsnamesm]{aastex63}

\usepackage{xspace}
\usepackage{amsmath}
\usepackage[multiple]{footmisc}
\shorttitle{GALLIFRAY}
\shortauthors{Saurabh et al.}

\graphicspath{{./}{figures/}}
\newcommand{\G}{\texttt{GALLIFRAY \xspace}}
\begin{document}
\title{GALLIFRAY -- A geometric modeling and parameter estimation framework for black hole images using bayesian techniques}


\author[0000-0001-7156-4848]{Saurabh}
\affiliation{P. D. Patel Institute of Applied Sciences, Charusat University, Anand, 388421 Gujarat, India}
\email{sbhkmr1999@gmail.com}
 
 \author[0000-0002-9608-1102]{Sourabh Nampalliwar}
\affiliation{Theoretical Astrophysics, IAAT, Eberhard Karls Universit\"{a}t, T\"{u}bingen, Germany}
\affiliation{SISSA, Via Bonomea 265, 34136 Trieste, Italy and INFN Sezione di Trieste, Italy}
\affiliation{IFPU - Institute for Fundamental Physics of the Universe, Via Beirut 2, 34014 Trieste, Italy}
\email{sourabh.nampalliwar@uni-tuebingen.de}

\begin{abstract}

Recent observations of the galactic centers of M87 and the Milky Way with the Event Horizon Telescope have ushered in a new era of black hole based tests of fundamental physics using very long baseline interferometry (VLBI). Being a nascent field, there are several different modeling and analysis approaches in vogue (e.g., geometric and physical models, visibility and closure amplitudes, agnostic and multimessenger priors). We present \texttt{GALLIFRAY}, an open-source Python-based framework for estimation/extraction of parameters using VLBI data. It is developed with modularity, efficiency, and adaptability as the primary objectives. This article outlines the design and usage of \texttt{GALLIFRAY}. As an illustration, we fit a geometric and a physical model to simulated datasets using markov chain monte carlo sampling and find good convergence of the posterior distribution. We conclude with an outline of further enhancements currently in development.

\end{abstract}

\section{Introduction} \label{sec:intro}

A supermassive black hole (SMBH) is thought to be the central engine at the core of most galaxies, where accreting gas propels powerful relativistic jets at large scales~\citep{2019ARA&A..57..467B}. The Event Horizon Telescope (EHT) has recently started probing these highly energetic regions of our universe. Equipped with an extremely high angular resolution ($\sim 25\mu as$), EHT provides an excellent opportunity to study black holes at horizon scales. Imaging and modeling the center of galaxies is one of the major focus of the EHT. 
As more data becomes available, there is a growing need for independent analysis frameworks, readily available to use.

Recent EHT observations of the M87 and Sgr A* SMBH~\citep{2019ApJ...875L...1E,2022ApJ...930L..12E} have provided us with new observables. The shadow images and more generally the VLBI data have initiated probes of the innermost accretion/jet flows near horizon scales, and tests of new theories~\citep{2018ApJ...863..148P, Kawashima2020AJE, Yuan2021ConstraintsOD, Gott2018ObservingTS, 2022arXiv221115906C}. 
There have been various studies in the literature that test for the possibility of other compact objects and modified gravity in strong gravity regimes, along with the estimation of physical parameters such as source size, mass, and spin, using gravitational waves~\citep{2017A&A...608A..60G, 2019PhRvD.100j4036A}, x-rays~\citep{2018PhRvL.120e1101C,2019arXiv190508012A}, astrometry~\citep{2020A&A...636L...5G}, shadows~\citep{2021CQGra..38uLT01V, 2021arXiv210801190N, 2020PhRvL.125n1104P, 2016CQGra..33l4001J, 2015ApJ...814..115P} and so on. In this context, EHT has provided new insights into the processes happening at the horizon scales~\citep{2022ApJ...930L..16E, 2019ApJ...875L...6E, 2022ApJ...930L..21B}. The revelation of the shadow structure at our galactic center has opened a plethora of questions as well, such as the short period time-variability, absence of jets, accretion process etc.~\cite{2022ApJ...930L..15E, 2022ApJ...930L..16E}. It has also highlighted the fact that the presence of the shadow does not confirm the presence of an event horizon. There are other compact objects which may provide similar results. These `black hole mimickers' have been studied in the literature from both theoretical as well as observational approach~\citep{2022ApJ...930L..17E, 2020PhRvL.125n1104P, 2022arXiv220801995S, 2022arXiv220507787V, 2020MNRAS.497..521O, 2018NatAs...2..585M}.

All tests of the kind described above involve estimating/extracting information about model parameters from real/synthetic datasets that observe/simulate some observables. Thus, this approach is strongly dependent on the initial models supplied to the estimator, which uses them as templates to best fit the experimental data, the observables through which the model parameters relate to the data, and analysis techniques that determine how models are to be fitted to data. A wide variety exists on each of these fronts. For instance, one could use a geometric (derived from simple shapes) or a physical model (derived by solving physical equations) for the shape of the shadow. Even within physical models, there is a wide range of models to choose from (e.g., semi-analytic RIAF, general relativistic magnetohydrodynamics (GRMHD) based standard and normal evolution (SANE) models, magnetised advection-dominated models (MAD) etc. ).

Moreover, it also depends on the choice of observables/data products and the likelihood function. The complex visibilities in radio astronomy are typically modeled using Gaussian distributions due to their well-understood statistical properties and suitability for model-fitting applications. However, there exists a diverse selection of other quantities that can be used instead of or in addition to the visibilities, such as closure phases and amplitudes, polarization fractions etc., and to analyse the data, various methods/techniques can be employed. Numerous codes are also present in literature to tackle specific challenges such as \texttt{Themis} \citep{Broderick_2020} which is a parameter estimation package used by
the EHT, \texttt{Comrade} \citep{Tiede2022} which is a \texttt{Julia} based open library aiming to produce VLBI images of black holes and active galactic nuclei, \texttt{eht-dmc} \citep{2021AJ....161..178P} which is a python Bayesian polarized imaging package etc.

With such a wide range of choices, and no single best choice, it is essential to establish a comprehensive framework that is able to incorporate as many different choices as possible, that can be used to gain insights not only about the particular system under consideration but about the larger physical reality. 

With this idea as the basis,
we present \texttt{GALLIFRAY}\footnote{\href{https://github.com/Relativist1/Gallifray}{https://github.com/Relativist1/Gallifray}} \citep{saurabh_sourabh_nampalliwar_2023}, a publicly available Python-based  parameter estimation framework that allows the users to fit their accreting black hole shadow model to radio interferometric observations (like EHT) using different techniques. At its core, the framework acts as a bridge between the user-defined models, observables, and likelihood functions on one hand, and the affine invariant MCMC ensemble sampler \texttt{emcee}\footnote{\href{https://github.com/dfm/emcee}{https://github.com/dfm/emcee}}~\citep{2013PASP..125..306F} on the other.

The rest of the paper is organized as follows. In Section \ref{sec:1}, we give an overview of the library and different implementations. In Section \ref{sec:observables}, we describe the observable under consideration. In Sections \ref{sec:2} and \ref{sec:3}, we describe the details of two models (geometrical and physical, respectively) that have been integrated in the library. The details pertaining to the Bayesian inference are explained in Section \ref{sec:4}. In Section \ref{sec:5}, we perform some simulated tests to illustrate the usage of the library, and we conclude and summarize in Section \ref{sec:6}.

\section{Gallifray modules} \label{sec:1}
\G consists of six different modules with different implementations based on the workflow. This includes
\begin{itemize}
    \item \textit{Models} : This module generates the model intensity/visibilities either from the ones already implemented (geometrical) or supplied by the user (physical). Geometric models are analytical models which are purely phenomenological in nature and do not arise from the physics of accretion or jets, but are based on the geometric properties. For raytracing models, users need to specify the running routine which could be in C/C++, fortran or python.
    
    \item \textit{Likelihood} : This module consists of the implementation of different types of posterior likelihood classes that could be based on the quantity to be compared.
    \item \textit{Priors} : This module consists of the information on the priors of the parameters on which the models are dependent. It further contains classes of different types of priors. This module can be further extended based on the usage of the user to include different priors than the ones that come by default. 
    \item \textit{Tardis} : \G comes with an inbuilt plotting module `Triangular Distribution Plotting (Tardis)' for visualizing the posterior distribution of estimated parameters.
\end{itemize}
A typical workflow of \G flows along the following steps:
\begin{enumerate}
    \item Define the observed/simulated datasets.
    \item Select a model to compare from the \textit{Models} module
    \item Define a tuple of initial guesses for the parameters of the model.
    \item Setting the appropriate priors from the \textit{Priors} module.
    \item Initialize the sampler by supplying the information set up in previous steps to create an instance, and then run the sampler for an appropriate number of iterations and walkers\footnote{refer to \texttt{emcee} documentation for a detailed overview on walkers}. Depending on the number of cores, users can set the number of tasks to run in parallel, when running the sampler.
    \item Once the sampler has computed the posterior distribution, plot the output data and extract the mean value of the parameters along with their respective error margins.
\end{enumerate}
\begin{figure*}[ht!]
    \centering
    \includegraphics[scale=0.45]{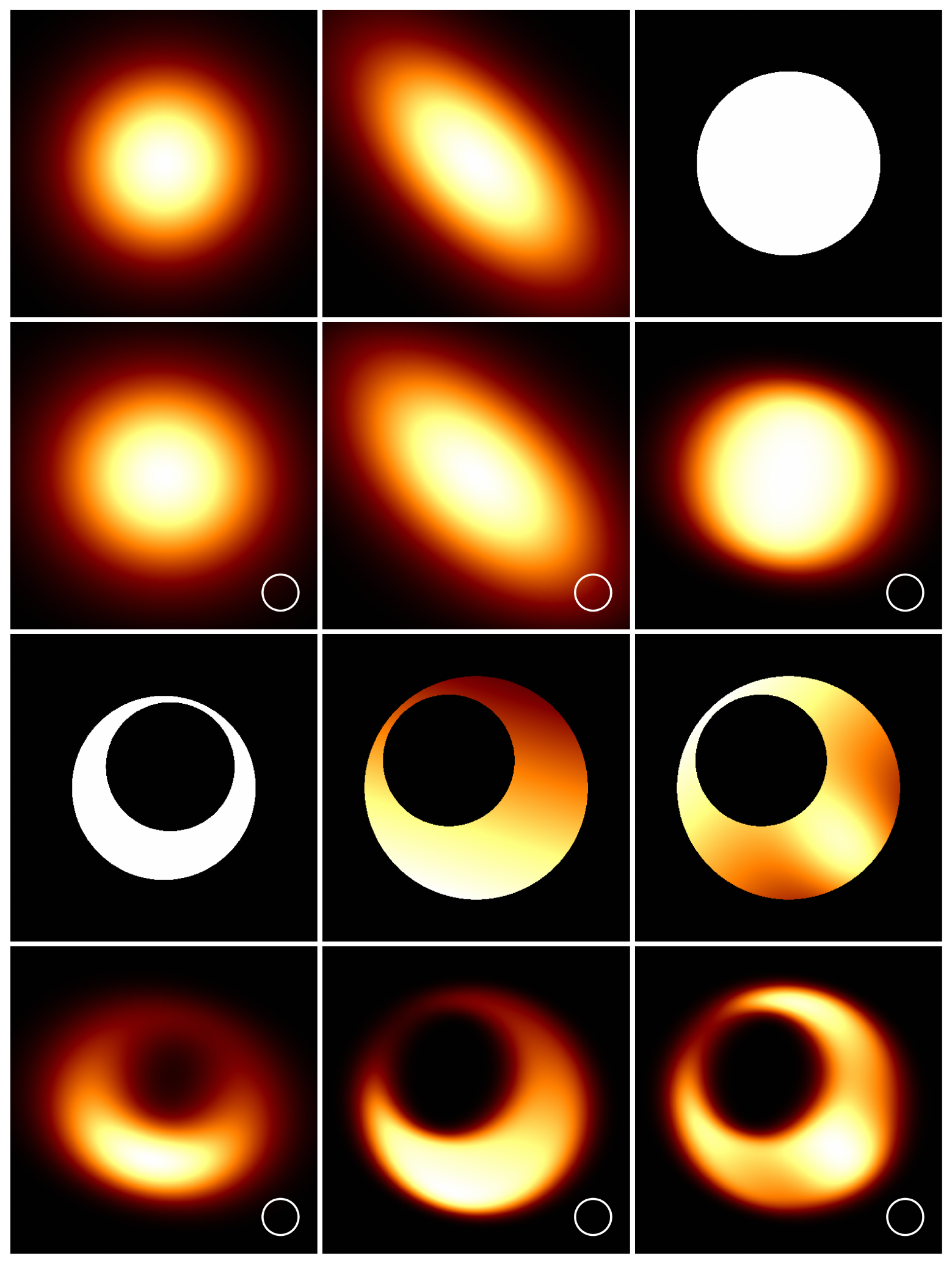}
    \caption{Sky domain images of the implemented models described in Section~\ref{sec:3}. The first and third rows are the unblurred models, while the second and fourth rows are blurred with the $50\%$ of the nominal beam size of the EHT array. The first row (left to right) corresponds to the symmetric Gaussian, asymmetric Gaussian, and Disk models respectively, and the third row corresponds to the Crescent, Xsring and Xsringauss model respectively.}
    \label{fig:models}
\end{figure*}

\section{Observables}\label{sec:observables}
There are various observables that can be constructed from VLBI observations, but the primary data product that any radio interferometer records are the complex visibilities $\mathcal{V}(u,v)$. Ideally, the intensity domain $I(x,y;\mathbf{\Theta_p})$ and the visibility domain is related to each other by the van Cittert-Zernike theorem~\citep{2017isra.book.....T}
\begin{equation}
    \mathcal{V}(u,v;\mathbf{\Theta_p}) = \int\int e^{-2\pi(xu+yv)}I(x,y;\mathbf{\Theta_p})dxdy.
\end{equation}
These visibilities are recorded over different baselines. These baselines correspond to the separations between the antennas in the interferometer, and their lengths and orientations determine the spatial scales and orientations that can be probed by the observations. Specifically, longer baselines correspond to higher spatial resolution, allowing finer details to be resolved in the observed images. Similarly, baselines oriented in different directions provide sensitivity to different features in the observed source. Therefore, by selecting a suitable set of baselines, radio interferometry can probe a wide range of spatial scales and orientations in the source structure. As the Earth rotates, the radio interferometer moves along a locus of points. Depending on the source's coordinates and observation time, these loci of points form different ellipses. Hence, an increase in the number of observed hours will result in more recorded visibilities and better sampling of the Fourier domain
leads to improved image reconstruction. The calibrated datasets of the EHT 2017 observations of Sgr A* and M87 (visibilites) are also publicly available \footnote{\href{https://github.com/eventhorizontelescope/2019-D01-01}{https://github.com/eventhorizontelescope/2019-D01-01}}\textsuperscript{,}\footnote{\href{https://github.com/eventhorizontelescope/2022-D02-01}{https://github.com/eventhorizontelescope/2022-D02-01}}.

In practice, other observables are also used for different purposes, such as closure phases, closure amplitudes, interferometric polarization fractions, etc.~\citep{2020ApJ...904..126B}. In this work, we utilize visibility amplitudes as the quantities to compare (although users will have the capability to introduce different quantities and construct their own likelihood functions). Since, the intensities and visibilities are Fourier pairs, it would be crucial to transform one domain to another. For this purpose, \texttt{GALLIFRAY} accepts Stokes-\textit{I} ($I(x,y;\mathbf{\Theta_p})$) values from the model and then generates the complex visibilities utilizing \texttt{ehtim}\footnote{\href{https://github.com/achael/eht-imaging}{https://github.com/achael/eht-imaging}}.
We will use this to construct a likelihood function ($\mathcal{\tilde{L}}(\mathbf{\Theta_{p}}| \textit{data})$) for our analyses. This library has been catered to handle \texttt{uvfits} format, particularly EHT-related datasets.

\section{Geometric Models} \label{sec:2}
Here, we review some of the geometric models that are currently available in \texttt{GALLIFRAY}. Note that these models do not arise from the modeling of physics of accretion or jets, rather, they are analytical models based on simple shapes. Owing to their simplicity, these models have been used and studied extensively in the past as a phenomenological tool to analyze data \citep{2008Natur.455...78D, 2019ApJ...875L...6E, 2022ApJ...930L..16E}.
\begin{figure}
    \centering
    \includegraphics[width=\columnwidth]{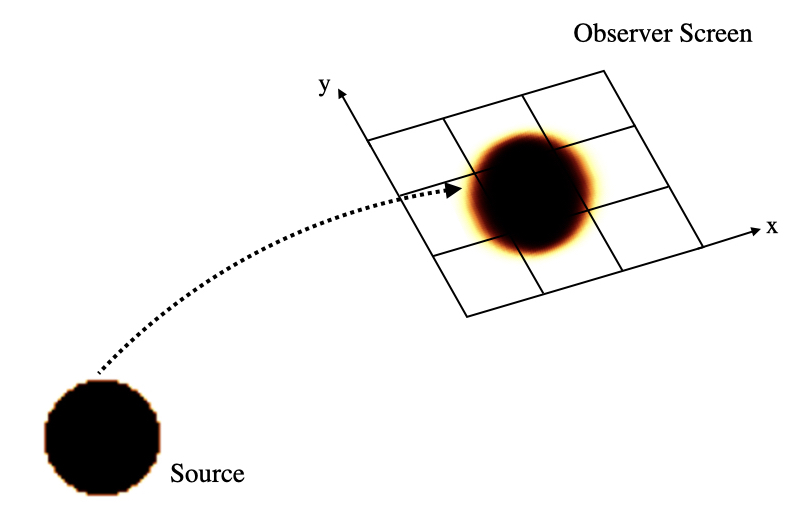}
    \caption{Schematic of the source and its image on the observer's screen.}
    \label{fig:schematic}
\end{figure}

As an example, the first and third row of Fig.~\ref{fig:models} represents the intensity distribution in the observer frame for the models described in the sections below, meanwhile the second and fourth row represents the intensity distribution convolved with $50\%$ of nominal beam size of the EHT configuration, for the images corresponding the same column.
\subsection{Symmetric Gaussian}
 One of the simplest geometric models employs a symmetric Gaussian for a radio source, and has been used extensively for model comparison in VLBI and radio interferometry for a long time
 ~\cite{2008Natur.455...78D}. 
This model gives the intensity on the observer's screen as shown in the schematic.~\ref{fig:schematic} as a 2-parameter function of a Gaussian, with the width $\sigma$ and the amplitude $V_0$ as the two parameters.
The corresponding function is given by
\begin{equation}
    I = V_{0}\exp{\left(-\frac{\alpha}{\sigma^2}\right)},
\end{equation}
where $\alpha$ is the grid coordinate in the angular form related to the position angle $\phi$.

\subsection{Asymmetric Gaussian}
The next simplest model is a generic Gaussian which inherits the property of asymmetry in the structure as well. Being more general than the symmetric Gaussian, this model provides better constraints. We use the modeling parameterizations used in~\cite{2008Natur.455...78D}. which is a 4-parameter model: the generalized width $\sigma$, anisotropy parameter $A$ (which introduces the asymmetry in the model), the position angle $\phi$ (for the orientation), and the intensity as the amplitude $V_{00}$. The corresponding function is given by 
\begin{equation}
    I = V_{00}\exp{\left(-\frac{(\alpha_{X} + \alpha_{Y})^2}{2\sigma_{min}^2}-\frac{(\alpha_{X} + \alpha_{Y})^2}{2\sigma_{maj}^2}\right)},
\end{equation}
where $\alpha_{X,Y}$ are the coordinates in angular form related to the position angle $\phi$ and $\sigma_{maj,min}$ are the widths of major and minor axis related to $\sigma$ and $A$ as 
\begin{eqnarray}
    \left(\sigma_{maj}, \sigma_{min}\right) = \left(\frac{\sigma}{\sqrt{1-A}},  \frac{\sigma}{\sqrt{1+A}}\right).
\end{eqnarray}

\subsection{Disk Model}
A simple, geometric disk model is also implemented in \texttt{GALLIFRAY}. The model is described by only two parameters, the overall size or radius, $\mathcal{R}$ and the intensity, $V_0$. In the 2-D intensity domain, this can be described as~\citep{2013MNRAS.434..765K},
\begin{equation}
\text{circ}\left(\frac{r}{\mathcal{R}}\right) = 
\begin{cases}
         V_0, &\text{if r }  \leq \mathcal{R} \\
        0, &\text{if r }  \geq \mathcal{R} \\ 
    \end{cases}.
\end{equation}
We can also find the visibilities analytically in the visibility domain. The Fourier transform of the disk is given by
\begin{equation}
    \mathcal{D}(u,v) = \frac{\mathcal{R}}{\sqrt{(u^2+v^2)}}J_1(2\pi\sqrt{(u^2+v^2)}\mathcal{R}),
\end{equation}
where $J_1(x)$ is the Bessel function of the first kind of the first order.
\subsection{Crescent and Generalised Crescent Models}
Here, we briefly discuss some of the physically motivated geometric models. These models are much more intuitive than the ones discussed above. They have been studied extensively in the light of EHT observations and also some early observations of Sagittarius A* using VLA.

\subsubsection{A Simple Crescent Model}
A circular crescent model can easily be constructed geometrically by simply subtracting two disks of different radii. The main feature here is that the inner disk can be shifted in any direction, giving it a crescent-like feature that can be seen in Fig.~\ref{fig:models}. In the intensity domain, this can be written as
\begin{equation}
    \text{Crescent}(r) = \text{circ}\left(\frac{r}{\mathcal{R}_{out}}\right) -  \text{circ}\left(\frac{r-d}{\mathcal{R}_{in}}\right),
\end{equation}
where $d$ is the distance between the centers of the disks and $\mathcal{R}_{out}>\mathcal{R}_{in}$. Following~\cite{2013MNRAS.434..765K}, we can write this in the visibility domain as
\begin{eqnarray}
    V_1 &=& 2\pi I_{0}\mathcal{D}_{out}(u,v; \mathcal{R}), \notag \\
    V_2 &=& 2\pi I_{0}e^{-2\pi i(au+bv)/\lambda}\mathcal{D}_{in}(u,v; \mathcal{R}), \notag \\
    V(u, v) &=& V_1 - V_2,
\end{eqnarray}
where $a$ and $b$ are the positions of the centers of both disks. This model can be parameterized for parameter estimation by introducing new parameters, the overall radius or size $R$; the relative thickness $\psi$, asymmetry parameter $\tau$; and the position angle or orientation $\phi$. Thus the final set of parameters are
\begin{equation}
    \mathbf{\Theta_{p}} = [I_0, \mathcal{R}_{out}, \psi, \tau, \phi],
\end{equation}
whose relation with the original parameters can be written as 
\begin{eqnarray}
    R &=& \mathcal{R}_{out} \notag \\
    \psi &=& 1 - {\mathcal{R}_{in}}/{\mathcal{R}_{out}} \notag \\
    \tau &=& 1 - {\sqrt{a^2+b^2}}/({\mathcal{R}_{out}-\mathcal{R}_{in}}) \notag \\
    \phi &=& \tan^{-1}(b/a) .
\end{eqnarray}
\subsubsection{Xsring Model}
\G also includes some more sophisticated models, which are the \textit{Xsring} and the \textit{Xsringauss} models discussed in~\cite{2016arXiv160900055B, 2019ApJ...875L...6E, 2022ApJ...930L..15E}. To construct the models, we first make a ring structure using the formalism introduced in the previous section. Then we introduce the ``slash" operation $\mathcal{S}(x)$ on the image to define a linear brightness gradient across the structure. In the intensity domain, this can be written as 
\begin{equation}
\label{eqn:10}
    \mathcal{S}(x) = \frac{h}{2} + x\left(\frac{h}{2\mathcal{R}_{out}}\right),
\end{equation} 
\\
where $h$ is the maximum intensity. As a result, the function will have the maximum intensity $h$ at the point $(\mathcal{R}_{out}, 0)$ and zero brightness at $(-\mathcal{R}_{out}, 0)$. We normalize the intensity to the unity, such that $h$ is
\begin{equation}\label{eqn:11}
    h = \frac{2}{\pi(\mathcal{R}^{2}_{out} - \mathcal{R}^{2}_{in})}.
\end{equation}

This introduces a non-linear brightness to the ring structure and provides a better fit to the data. This function is then simply applied to the crescent structure (discussed in the previous section),
\begin{equation}
    I(x,y) = \mathcal{S}(x)(\mathcal{D}_{out}(x,y) - \mathcal{D}_{in}(x,y)).
\end{equation}
The corresponding function in the visibility domain can be written as
\begin{equation}\label{eqn:13}
    V(u,v) = \frac{h}{2}\left(\text{circ}(\rho, \mathcal{R}) + \frac{i}{2\pi}\frac{d}{d\rho}\text{circ}(\rho, \mathcal{R})u\right),
\end{equation}
where
\begin{eqnarray}
    \frac{d}{d\rho}\text{circ}(\rho, \mathcal{R}) &=& \mathcal{R}\bigg[ \frac{\pi \mathcal{R}\left(J_0(2\pi\mathcal{R}\rho) - J_2(2\pi\mathcal{R}\rho)\right)}{\rho^2} \\ &-&  \frac{J_1(2\pi\mathcal{R}\rho)}{\rho^3} \bigg], \nonumber
\end{eqnarray}
where $J_{n}(x)$ is the Bessel function of the first kind of order $n$. Therefore the final version of the model is
\begin{equation}
\label{eqn:15}
    V(u, v) = V_{out}(u, v) - V_{in}(u, v),
\end{equation}
where $V_{out/in}$ is the visibility for the outer/inner disk using Eq.~\ref{eqn:13}. After this, we can further allow the model to rotate by some position angle $\phi$.
Then the final set of parameters for estimation are, $I_0$ (intensity/amplitude of the disk), $\mathcal{R}_{out}$ (outer disk radius),$\epsilon$ (eccentricity), $f$ (fading parameter), $\phi$ (position angle),
\begin{equation}
    \mathbf{\Theta_{p}} = [I_0, \mathcal{R}_{out}, r_q, \epsilon, f, \phi].
\end{equation}
Here $r_q={\mathcal{R}_{out}}/{\mathcal{R}_{in}}$, $f=h_1/h_2$, $\epsilon={d}/({\mathcal{R}_{out}-\mathcal{R}_{in}})$ where $d$ is the distance between the disks. When $f=0$ the brightness of the ring grows from zero to unity, while when $f=1$, the brightness is uniform.

\subsubsection{Xsringauss Model}
This model is similar to the Xsring model but with an additional emission component, an elliptic Gaussian.
We can define the maximum brightness of the model in a similar way as Eq.~\ref{eqn:11}
\begin{eqnarray}
    h &=& \frac{2}{\pi}\bigg[\left(\mathcal{R}^2_{out} - \mathcal{R}^2_{in}\left(1+\frac{d}{\mathcal{R}_{out}} \right) \right)f \\ &+& \left(\mathcal{R}^2_{out} -  \mathcal{R}^2_{in}\left(1+\frac{d}{\mathcal{R}_{out}} \right) \right) \bigg]^{-1}. \nonumber
\end{eqnarray}
The slash operation then can be defined in a similar way as Eq.~\ref{eqn:10}:
\begin{equation}
    \mathcal{S}(x) = \frac{(h_1 + h_2)}{2}+\frac{(h_1 - h_2)}{2}\left(\frac{x}{\mathcal{R}_{out}}\right),
\end{equation}
where $h_1$ and $h_2$ are the intensities at $x=\mathcal{R}_{out}$ and $x=-\mathcal{R}_{out}$. The corresponding visibility function for the model will be
\begin{eqnarray}
    V_{out}(u, v) &=& \xi \bigg( \text{circ}(\rho, \mathcal{R}_{out})\\ &+& \frac{1}{2\pi}\left( \frac{\eta}{\xi}\right)\frac{d}{d\rho}\text{circ}(\rho, \mathcal{R}_{out})u \bigg) \nonumber,
\end{eqnarray}
where $\xi=(h_1 + h_2)/2$ and $\eta=(h_1-h_2)/2$.

Using the analogy of Eq.~\ref{eqn:15}, we get the slashed ring visibility function for this model to be
\begin{equation}
    V_{sr}(u, v) = V_{out}(u, v) - e^{i2\pi u}V_{in}(u, v).
\end{equation}
Then an additional Gaussian is added. The Gaussian component causes the brightness to fall off more smoothly beyond the slashed ring. It is centered at the edge of the inner ring, at $x=\mathcal{R}_{in}-d$, giving a shift factor in the Fourier domain as $e^{i2\pi u(\mathcal{R}_{in}-d)}$ resulting the visibility to be 
\begin{equation}
    V_{\text{gauss}}(u, v) = e^{-2\pi^2k^2((u\alpha)^2 + (u\beta)^2)} e^{-i2\pi u(\mathcal{R}_{in}-d)}.
\end{equation}
where $k={1}/{2\sqrt{2\ln{2}}}$ is the coefficient, changing the FWHMs $(\alpha, \beta)$ into widths $(\sigma_{maj}, \sigma_{min})$. 
The final visibility of the Xsringauss model is given by
\begin{equation}
    V_{u, v} = I_0((1-g_{q})V_{sr} + g_{q}V_{\text{gauss}}).
\end{equation}
The additional elliptical Gaussian brightness is specified with three parameters: $g_{ax}$, $a_q$, and $g_q$. Main axis of the FWHM is expressed by $g_{ax}=\alpha/\mathcal{R}_{out}$, axial ratio is $a_q=\beta/\alpha$ and the fractional Gaussian flux $g_q$. When $g_q=0$, model is without Gaussian brightness and when $g_q=1$, means that the brightness component is entirely due to the Gaussian component.
Thus this nine-parameter models can be expressed via the given set of parameters
\begin{equation}
    \mathbf{\Theta_{p}} = [I_0, \mathcal{R}_{out}, r_q, \epsilon, f,g_{ax}, a_q, g_q, \phi].
\end{equation}
For analytical visibility domain functions of both these models, refer to~\cite{2016arXiv160900055B, 2019ApJ...875L...6E}.
\section{Physical Models} \label{sec:3}
In the previous section, we discussed models that were motivated geometrically and constructed without considering any physical scenarios. In reality, we would want to constrain a number of physical parameters, such as the mass of the central object $(M)$, spin parameter $(a^*)$, etc. Furthermore, there will be magnetic fields  generated by the accretion disk, along with several other details about the plasma. To model these properties and compute the radiation from the accretion disk general relativistic raytracing and radiative transfer or GRRT is employed, wherein the geodesic equations are solved for a given spacetime and then the intensity is calculated by solving the relativistic radiative transfer equations. Various numerical codes and schemes have been studied and proposed in the literature such as \texttt{ipole} \citep{2018MNRAS.475...43M}, \texttt{RAPTOR} \citep{2018A&A...613A...2B}, \texttt{BHOSS} \citep{younsi_porth_mizuno_fromm_olivares_2020}, \texttt{RAIKOU} \citep{2021arXiv210805131K}, \texttt{Odyssey} \citep{2016ApJ...820..105P}, \texttt{grmonty} \citep{2009ApJS..184..387D}, \texttt{grtrans} \citep{2016MNRAS.462..115D}, \texttt{Arcmancer} \citep{2018ApJ...863....8P}, \texttt{GRay} \citep{2013ApJ...777...13C}, \texttt{ARTIST} \citep{2017MNRAS.464.4567T} etc. Refer to \cite{2020ApJ...897..148G} for a detailed comparative analysis of these codes.
\begin{figure*}
    \centering
    \includegraphics[width=\textwidth]{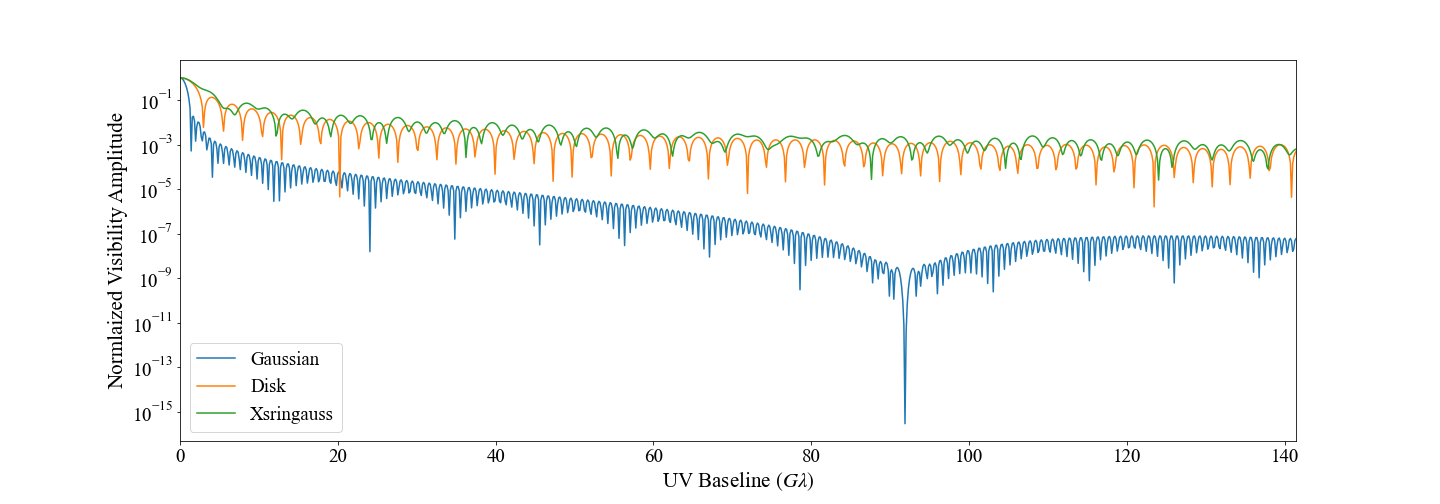}
    \caption{Logarithmic normalised visibility amplitudes for the models discussed in Section~\ref{sec:2}}.
    \label{fig:vis_amp}
\end{figure*}

\subsection{RIAF Model} \label{riaf}
In this section, we briefly discuss the implementation of a publicly available general relativistic, polarized radiative transfer code \texttt{ipole}~\citep{2018MNRAS.475...43M}. It is written in standard \texttt{C} language. We use ipole as an example of a raytracing code used in \texttt{GALLIFRAY}. The public version consist an implementation of a semi-analytic RIAF (Radiatively Inefficient Accretion Flows) model~\citep{2006MNRAS.367..905B,2018ApJ...863..148P} which utilizes fitting formulas for emission and absorption coefficients \citep{2016ApJ...822...34P} to solve the polarised general relativistic radiative transfer equation.

In this work, we use thermal distribution for synchrotron emission and restrict to just Stokes-\textit{I} parameter. Here we use a single population of thermal electrons orbiting a Kerr black hole with a toroidal magnetic field~\citep{2018ApJ...863..148P}. The number density and temperature of electrons is given by
\begin{eqnarray}
    n_{th} &=& n_{e}r^{-\alpha}e^{-z^2/2h^2}, \\
    T_{th} &=& T_{r}r^{-\beta}, 
\end{eqnarray}
where $z=r\cos\theta$ and $h$ is the disk width. The magnetic field strength of the toroidal field is then given by
\begin{eqnarray}
    \frac{B ^ 2}{8\pi} = \frac{1}{10}n_{th}\frac{m_p c^2}{6r}.
\end{eqnarray}
The final set of parameters are then, 
\begin{eqnarray}
    \mathbf{\Theta_{p}} = [a^*, \theta_{inc}, \alpha ,\beta, n_{e}, T_{e}],
\end{eqnarray}
where $\theta_{inc}$ is the inclination angle. For the analysis, we keep the model simple for comparison by fixing some of the parameters such as the spectral index $\alpha=1.5$ and $\beta=0.84$. This sample dataset is adopted from the Themis analysis~\citep{Broderick_2020} for the benchmark test.
To perform MCMC for a raytracing model, the package comes with a template that instructs how one can modify their raytracing code to be integrated with \texttt{GALLIFRAY}.

\section{Parameter Estimation}\label{sec:4}
Once, we are equipped with the model template, we can now proceed to recover the parameters. There are various methods in the literature that are used in practice, such as the least squares method, maximum likelihood estimation, principal component analysis, etc., for estimation. We follow the MCMC approach for our purpose.
For a set of parameters for a given model, we can define the 'likelihood' $\mathcal{\tilde{L}}(\mathbf{\Theta_{p}}| \textit{data})$ given by the standard Bayes's theorem,
\begin{eqnarray}
    P(\mathbf{\Theta_{p}}| \textit{data}) = P_{pr}(\mathbf{\Theta_{p}})\mathcal{\tilde{L}}(\mathbf{\Theta_{p}}| \textit{data}),
\end{eqnarray}
where $P_{pr}(\mathbf{\Theta_{p}})$ and $P(\mathbf{\Theta_{p}}| \textit{data})$ are the prior and posterior distributions.
For the implementation of the algorithm, we use the \texttt{emcee}~\citep{emcee} package, and integrate it with the library models.

\subsection{Priors}
\G comes with some pre-defined priors for the geometric models. 
Currently, this includes:
\begin{enumerate}
    \item uniform/flat : requires two boundary values,
    \item Gaussian : requires a mean and standard deviation value.
\end{enumerate}

The flexibility of the library gives an user the freedom to prompt their own priors.
\subsection{Likelihood}\label{sec:4.2}
For constructing the likelihood function, we first need to define the observational quantities that will be compared (e.g., visibility amplitudes, phases, closure amplitudes, and closure phases). \G has the flexibility to work with any of these observational quantities. Currently, it includes visibility amplitudes and the following likelihood functions based on visibility amplitudes:

\subsubsection{Gaussian Visibility Amplitudes}

\G comes with log-likelihood, which assumes Gaussian errors for visibility amplitudes~\citep{2017isra.book.....T}:
\begin{eqnarray}
    \mathcal{L}(\mathbf{\Theta_{p}}) = -\sum_{i} \frac{(V_{i} - \tilde{V}_i)^2}{2\sigma_j^2},
\end{eqnarray}
where $V_{i}$ and $\sigma_j$ are the observed visibility amplitudes and corresponding errors, and $\tilde{V}_i$ are the visibility amplitudes of the model supplied.
\subsubsection{Norm-marginalised Log-likelihood}
We define the norm-marginalised log-likelihood as defined in~\cite{2014ApJ...784....7B}.
\begin{eqnarray}
     \mathcal{\tilde{L}} = \mathcal{L}_{max} + \frac{1}{2}\log\left( \frac{2\pi V^2_{0,max}}{\sum_i \tilde{V}_i^2/\sigma_i^2} \right) ,
\end{eqnarray}
where
\begin{eqnarray}
    \mathcal{L}_{max} &=& -\frac{\left(\sum_i V_j^2/\sigma_i^2\right)\left(\sum_i \tilde{V}_i^2/\sigma_i^2\right) - \left(\sum_i V_i\tilde{V}_i/\sigma_i^2\right)}{2\sum_i \tilde{V}_i^2/\sigma_i^2}, \notag \\
    V_{0,max} &=& \frac{\sum_i V_i\tilde{V}_i/\sigma_i^2}{\sum_i \tilde{V}_i^2/\sigma_i^2}.\notag
\end{eqnarray}

Here, $V_{i}$ and $\sigma_j$ are the observed visibility amplitudes and corresponding errors, and $\tilde{V}_i$ are the model visibility amplitudes.
\begin{figure}
    \centering
    \includegraphics[width=\columnwidth]{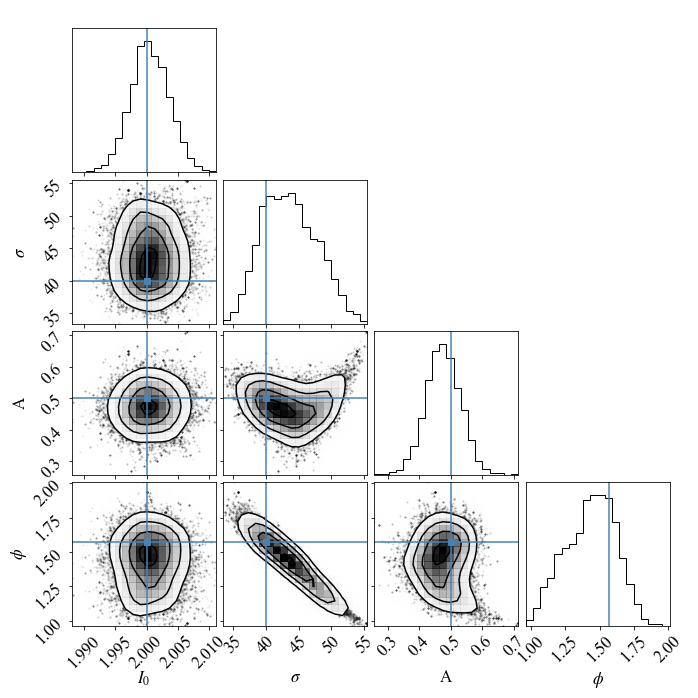}
    \caption{Posterior distribution for asymmetric Gaussian model discussed in Section~\ref{sec:5}.}
    \label{fig:gauss_mcmc}
\end{figure}

\begin{figure}
\centering
    \includegraphics[width=\columnwidth]{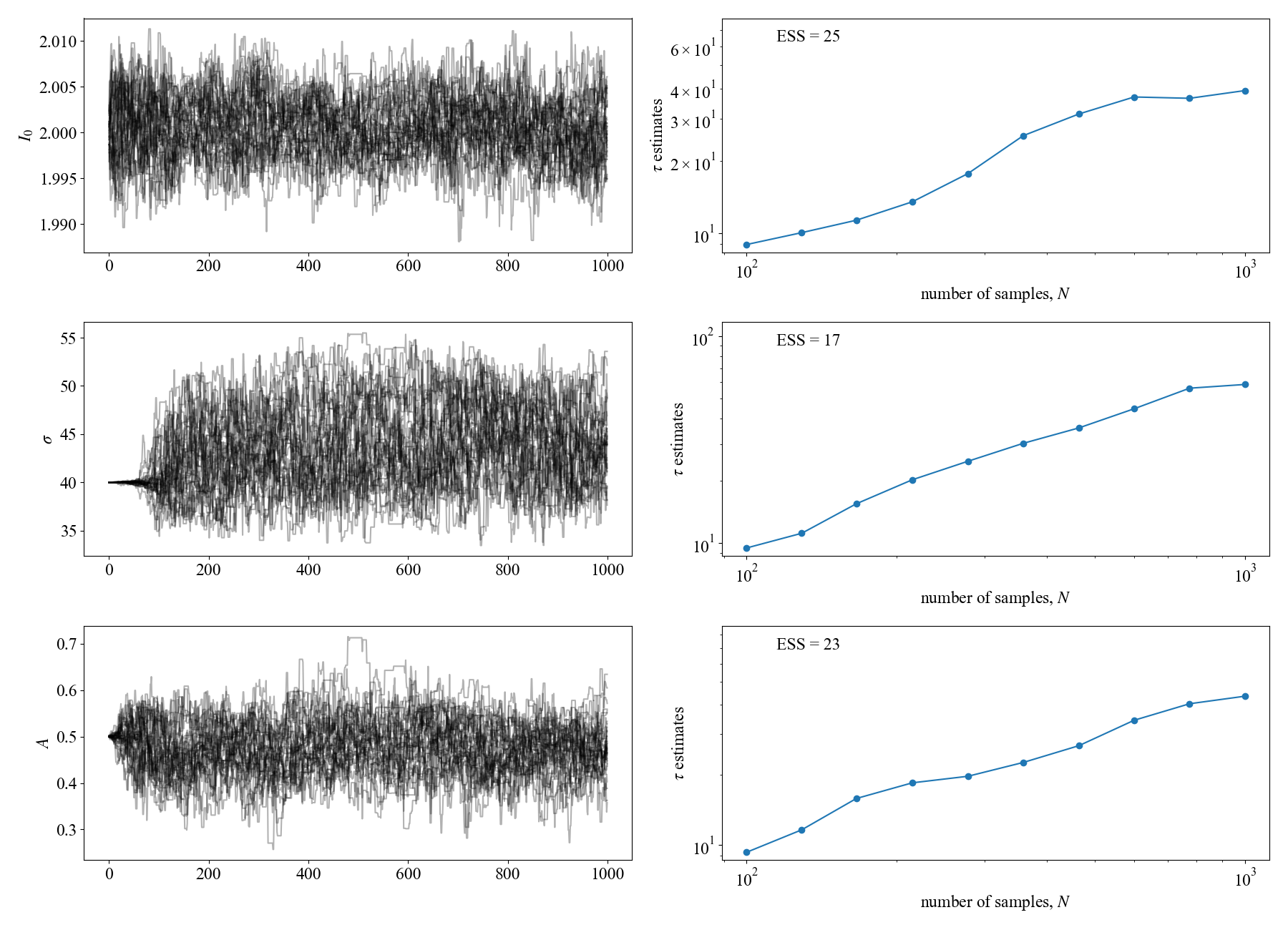}
    \caption{Trace plot and the corresponding autocorrelation time plot for the asymmetric Gaussian model discussed in Section~\ref{sec:gauss}.}
    \label{fig:gauss}
\end{figure}

\begin{figure}
    \centering
    \includegraphics[width=\columnwidth]{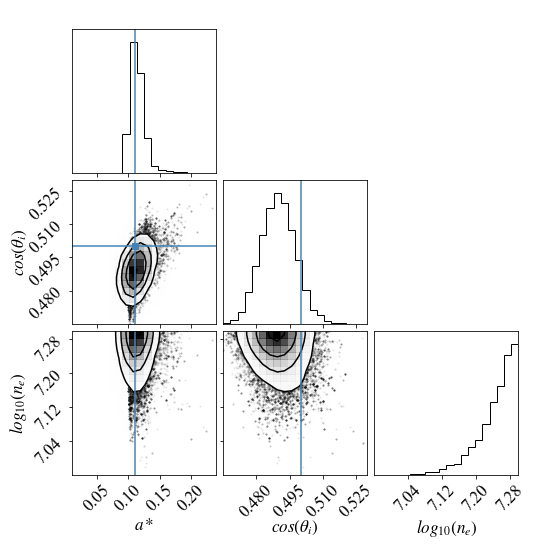}
    \caption{Posterior distribution for RIAF fit of ipole model discussed in Section~\ref{sec:5}.}
    \label{fig:ipole_fit}
\end{figure}

\begin{figure}
 \includegraphics[width=\columnwidth]{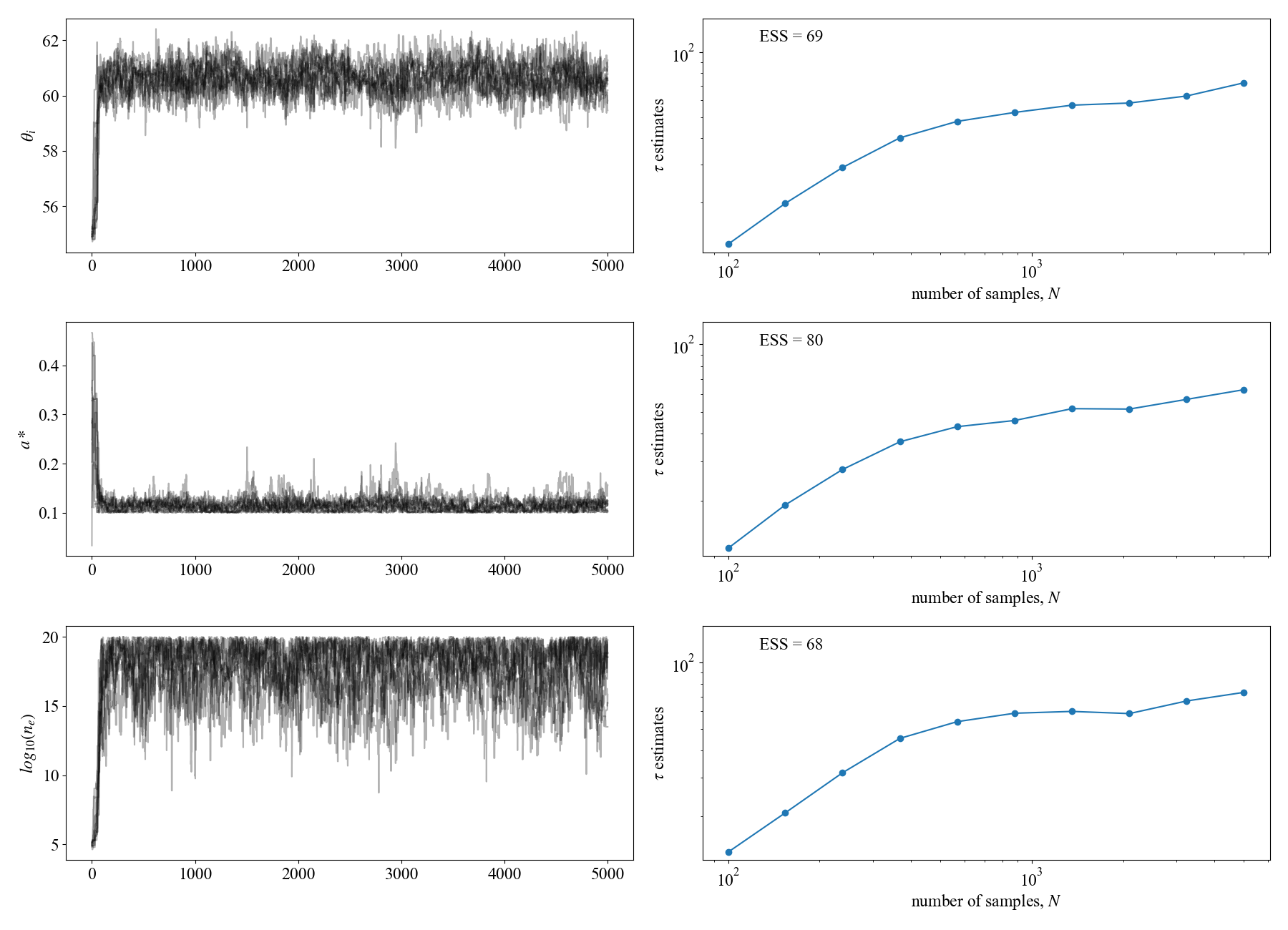}
    \caption{Same as Fig.~\ref{fig:gauss} for \texttt{ipole} raytraced model discussed in Section~\ref{sec:riaf_ipole}.}
    \label{fig:ipole}
    \end{figure}
\section{Simulated Tests}\label{sec:5}
Here, we demonstrate the capability of the library to do data analysis and parameter estimation using simulated VLBI data sets. We use \texttt{ehtim}~\citep{2018ApJ...857...23C} library to manipulate and simulate VLBI datasets by introducing geometric and raytraced models as the reference models. We present two different analyses to corroborate the accuracy of the sampler and estimate the parameters. To demonstrate the accurate working of the models, we first simulate the visibility amplitudes of some of the models. Fig.~\ref{fig:vis_amp} shows the corresponding logarithmic visibility amplitudes over baseline length.

To generate synthetic datasets, we utilized the EHT 2017 observation campaign with 24-hour observation, which indicated the most optimistic view. To model a more realistic scenario, we include thermal noise in the VLBI datasets as well. In this way, we can mimic, to the best of the ability, an observation campaign.
We refer to a set of 4 synthetic datasets: 2 geometric and 2 raytraced models.
For geometric fitting, the asymmetric Gaussian and Xsring models have been used. For raytracing, we utilize the RIAF model mentioned in section \ref{riaf}.
\subsection{Asymmetric Gaussian}
\label{sec:gauss}
The first test is performed with the asymmetric Gaussian model. We generate a synthetic dataset for the given set of parameters: $ \mathbf{\Theta_{p}} = [I_0, \sigma, A, \phi] = [2, 40, 0.5, 90^\circ]$, for 24 hours observation, considering the EHT 2017 array configuration.

We adopted 32 walkers and 1000 samples for the MCMC, while taking the truth values as the initial guesses, which ran for 400 core hours. The resultant posterior distribution is plotted in ~Fig. \ref{fig:gauss_mcmc} showing good convergence of the parameters with a burn-in of $1000$ samples. To show the convergence, we also show the trace plot and the corresponding autocorrelation plot as well. We plot the autocorrelation time ($\tau$) with the number of samples. Fig.~\ref{fig:gauss} is plotted for the Gaussian model. The effective sample size (ESS) is mentioned for each parameter.

\subsection{Semi-analytic RIAF}
\label{sec:riaf_ipole}
The second test is performed using a raytracing model. There are several general relativistic raytracing and radiative transfer codes in literature that solve unpolarised and/or polarised radiative transfer equations; see for eg.~\cite{2018A&A...613A...2B, 2016ApJ...820..105P, 2011CQGra..28v5011V, 2009ApJS..184..387D,2021arXiv210805131K}. In principle, any of the publicly available or stand-alone schemes can be used. For this purpose, we use \texttt{ipole}, which solves polarised radiative transfer equation in Kerr spacetime. It is primarily used to raytrace GRMHD datasets, but a semi-analytic scheme is also implemented for RIAF modeling. We utilize this model as an example and a test for the extraction of parameters. The model parameters (for synthetic dataset) include: $ \mathbf{\Theta_{p}} = [a^*, \theta_{i}, n_{e,th}] = [0.1,60^\circ,3.5 \times 10^7]$. This model can be extended to include other parameters as well, such as the position angle, non-thermal population of electrons, temperature normalization, etc. For simplicity, we ignore these and utilize just the above parameters.

We adopted a total of 50 walkers and 5000 samples for the MCMC sampling, which ran for 2200 core hours. The initial positions were kept as the initial guess. The posterior distribution is plotted in Fig.~\ref{fig:ipole_fit} with a burn-in of $1000$ samples. The inclusion of electron number density makes the case a bit complicated, as small variations in $n_e$ does not change the total flux and accretion by much. The parameter has a relatively high magnitude, and this would inherently mean that the parameter can occupy a large sample space. For this reason, the logarithm value of the parameter is used. Fig.~\ref{fig:ipole} shows the autocorrelation plot for the MCMC run along with the ESS, showing a clear trend and independence of the parameters as the chain length is increased. Increasing the number of samples shall provide us with more clarity but will not change the trend too much. It also verifies the previous analyses done in literature such as \cite{Broderick_2020}, from where we adopted the simulated values for the raytracing test. However, the reference dataset was taken to be the visibility amplitudes of Sgr A* from 2007 and 2009 \citep{2011ApJ...735..110B}. The spin and inclination parameter are constrained well as compared to the $n_e$ parameter. 

\section{Discussion and Summary} \label{sec:6}
We present an open-source python based library, \texttt{GALLIFRAY}, for parameter estimation and modelling of VLBI sources primarily on horizon scales. The framework is modular in nature and amenable to incorporating user-supplied models easily. At present, it includes multiple geometric models of the source (Gaussians, Disk, Crescent, Xsring, Xsringauss) described in Section~\ref{sec:3} and one physical model based on \texttt{ipole} described in Section~\ref{sec:4}.  The library uses the MCMC sampling with multiple likelihood functions: Gaussian visibility amplitudes and norm-marginalised log-likelihood described in Section~\ref{sec:4.2} and priors for parameter estimation and further analyses. It comes with various features such as implementation of different likelihood functions, priors and different modelling tools for the sources. At present, the library uses visibility amplitudes for the analyses. 



To illustrate the working of the library, we present two analyses: in the first analysis, we initially used an asymmetric Gaussian model as the source to model the intensity distribution and then generate synthetic dataset using the \texttt{ehtim} library taking considering thermal noise as well as phase errors into account. Afterwards, we employ the MCMC sampler by first defining the priors, Gaussian likelihood and other parameters (walkers, samples and initial guesses). As a result, we get the posterior distribution shown in Fig.~\ref{fig:gauss_mcmc}), delivering good convergence of recovered parameters.

In the second analysis, we focus on RIAF model simulated using \texttt{ipole}. We employ similar steps as done in the first analysis but with a physical model and also find good convergence of the parameters as shown in Fig.~\ref{fig:ipole_fit}.

This work marks the first version of the library. Upgrades from EHT to ngEHT will open a new era for studying more exotic and unexplored sources at very high angular resolutions, which will lead to more diversity in models, observables and analyses. To perform more rigorous tasks and increase the capability of the library, development along several fronts is ongoing:

\begin{itemize}
    \item Supermassive black hole binaries has intrigued a lot of interest in recent years, and the observations through VLBI can help us probe 
 extremely large structures with high precision. There are plans currently in motion to include and implement orbiting binaries for modeling supermassive black hole binaries~\citep{2022ApJ...927...93F}. This model can be identified as an astrometric binary model to estimate orbital parameters. This allows us to model two Gaussian sources on a Keplerian orbit without any general relativistic effects such as lensing, redshift, jets, flares etc.
    \item In the next phase, there will be several other likelihood functions which will be readily available for the user to do model fitting. These include the closure quantities which are currently being investigated in more detail.
\end{itemize}

Meanwhile, there are some long-term goals as well which requires more study and investigation. These are the implementations which may occur in the future versions of the library.
\begin{itemize}
    \item More recently, there have been other parameterizations for the geometric models that can model the source in a more sophisticated way. This includes the 'mG-ring Model', developed in \cite{2020SciA....6.1310J} and used in \cite{2022ApJ...930L..15E}.
    \item Moreover, with the extraordinary results of the EHT observations, it has also paved a secure way towards horizon-scale imaging with space-based VLBI. There have been several proposals to do sub-mm interferometry in space with the usage of multiple satellites on circular orbits around Earth~\citep{2013ARep...57..153K, 2022cosp...44.2068R, 2021A&A...650A..56R, 2021A&A...649A.116F, 2021MNRAS.500.4866A, 2021ApJ...922L..28J}. Adding the capability for creating synthetic datasets for space-based observations could be one of the implementations which the library can see in the future.
\end{itemize}


\section*{Acknowledgements}
S.N. acknowledges support from the Alexander von Humboldt Foundation and the Deutscher Akademischer Austauschdienst.

\bibliographystyle{aasjournal}
\bibliography{ref}

\end{document}